\documentclass[aps,pra,twocolumn,american,showkeys,showpacs,floatfix,reprint]{revtex4-1}
\usepackage{lmodern}
\usepackage[T1]{fontenc}
\usepackage[koi8-r,latin9]{inputenc}
\usepackage{geometry}
\usepackage{units}
\usepackage{amsmath}
\usepackage{amssymb}
\usepackage{graphicx}
\usepackage{setspace}

\makeatletter
 
 \@ifundefined{textcolor}{}
 {%
   \definecolor{BLACK}{gray}{0}
   \definecolor{WHITE}{gray}{1}
   \definecolor{RED}{rgb}{1,0,0}
   \definecolor{GREEN}{rgb}{0,1,0}
   \definecolor{BLUE}{rgb}{0,0,1}
   \definecolor{CYAN}{cmyk}{1,0,0,0}
   \definecolor{MAGENTA}{cmyk}{0,1,0,0}
   \definecolor{YELLOW}{cmyk}{0,0,1,0}
 }

\makeatother

\usepackage{babel}
\begin{document}

\title{Generalized binomial distribution in photon statistics}

\author{Aleksey V. Ilyin}

\email{a.v.ilyin@mail.mipt.ru}

\affiliation{Moscow Institute for Physics and Technology}

\date{\today}
\begin{abstract}
The photon-number distribution between two parts of a given volume
is found for an arbitrary photon statistics. This problem is related
to the interaction of a light beam with a macroscopic device, for
example a diaphragm, that separates the photon flux into two parts
with known probabilities. To solve this problem, a Generalized Binomial
Distribution (GBD) is derived that is applicable to an arbitrary photon
statistics satisfying probability convolution equations. It is shown
that if photons obey Poisson statistics then the GBD is reduced to
the ordinary binomial distribution, whereas in the case of Bose-Einstein
statistics the GBD is reduced to the Polya distribution. In this case,
the photon spatial distribution depends on the phase-space volume
occupied by the photons. This result involves a photon bunching effect,
or collective behavior of photons that sharply differs from the behavior
of classical particles. It is shown that the photon bunching effect
looks similar to the quantum interference effect. 
\end{abstract}



\maketitle

\section{Introduction\label{sec:Introduction}}

This article examines the spatial distribution of photons in a light
beam with arbitrary photon statistics. If the photons obey the Bose-Einstein
(BE) statistics, then the photon spatial distribution is found to
exhibit certain features, termed here the 'photon bunching effect',
or collective photon behavior that differs sharply from the behavior
of classical particles.

In the literature, the term \emph{photon bunching} is sometimes related
to the Brown-Twiss effect, which is explained by intensity fluctuations
in the light beam~\cite{Brown_Twiss}. In this paper, the term \emph{photon
bunching} is used in an entirely different sense, in that the effects
considered in this article are not associated with intensity fluctuations.
As shown in Sections \ref{sec:GBD_in_BE}-\ref{sec:Equivalence},
the photon bunching effect in the BE statistics is rather similar
to the quantum interference effect, which was first observed in~\cite{HOM_effect}.

The problem considered in this work is stated in Section~\ref{sec:Diaphragm},
which also contains the basic formula for the flux of classical particles
obeying Poisson statistics and for the flux of quantum particles obeying
BE statistics.

To solve the stated problem, a \emph{Generalized Binomial Distribution}
(GBD) is derived in Section~\ref{sec:G_B_D} for an arbitrary photon
statistics. It is shown that in the case of BE statistics, the GBD
reduces to the Polya distribution, which predicts the photon bunching
effect that is discussed in Section~\ref{sec:GBD_in_BE}.

The equivalence of various statistical problems is discussed in Section~\ref{sec:Equivalence},
suggesting the possibility of applying the results obtained to other
equivalent problems involving the interaction of photon flux with
a beamsplitter, photodetector, or neutral filter. The main conclusions
of this work are summarized in Section~\ref{sec:Conclusions}.

A theoretical approach developed in this paper allows one to study
the subtle features of spatial distribution of particles in the BE
statistics, as well as in any other photon statistics. In the limit
of large phase-space volumes, the results obtained in this work coincide
with the known classical solution, while in the limit of small phase
volumes the results are found to be consistent with an experiment~\cite{Stability_of_clusters}
which examined the phenomenon of quantum interference between two
photons incident on the \emph{same} input port of a beamsplitter.

\section{\label{sec:Diaphragm}Interaction of a light beam with a diaphragm
from the viewpoint of photon statistics}

\subsection{\label{sub:Statement}Statement of the problem}

\begin{figure}[ht]
\noindent \begin{centering}
\includegraphics[width=5cm]{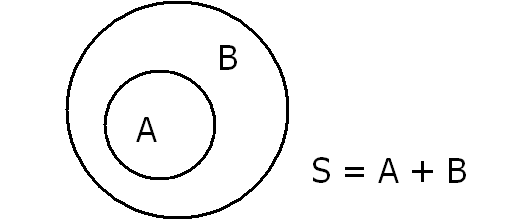} 
\par\end{centering}

\caption{{\small An illustration of a light beam cross-section, representing
the interaction of photon flux with a diaphragm when $k$ photons
pass through the hole $A$ while $n-k$ photons are absorbed by the
screen~$B$. Sampling volumes $A$ and $S$ may correspond to possible
detector apertures in a photon statistics measurement.} {\label{fig:Problem_statement}}}
\end{figure}

Consider a beam of light that has a uniform distribution of statistical
properties over its cross-section. Let the beam cross-section $s$
be divided into two \emph{arbitrary} parts, $a$ and~$b$. The photon
statistics in part~$a$ is measured during a certain sampling time~$\tau$,
\emph{i.\,e.} in sampling volume $A=ac\tau$, where $c$ is the speed
of light. Capital characters $A$, $B$, and $S$ will be used in
this paper to designate sampling volumes corresponding to light beam
cross-section areas $a$, $b$ and~$s$ (Fig.~\ref{fig:Problem_statement}).

If a sampling volume is normalized to a coherence volume, then it
will correspond to a certain phase-space volume, because a coherence
volume corresponds to a single phase-space cell. Therefore, volumes
of $A$, $B$ and $S=A+B$ may be considered as dimensionless phase-space
volumes consisting of an arbitrary number of cells. The term 'phase-space
cell' (or a bin) is used in this article as an equivalent to 'coherence
volume', which is the same as 'single mode of radiation'. 

Since the light beam is supposed to have a uniform distribution of
statistical properties over its cross-section, probabilities $\alpha$
and $\beta$ for a photon to be in volume $A$ or $B$, respectively,
are given by 
\begin{equation}
\alpha=\frac{A}{A+B},\,\,\,\,\,\,\,\,\,\,\,\,\,\beta=\frac{B}{A+B},\label{eq:alpha_beta}
\end{equation}
 so that $\alpha+\beta=1$.

Photon statistics in an arbitrary volume will be discussed in this
article. In volume $A$, the photon statistics is represented by an
infinite series $p_{n}(A)$, where $n=0,\,1,\,2,\,\ldots$ and $p_{n}(A)$
is the probability that $n$ photons are in volume~$A$.

The following problem is considered in this work:

Assume the photon statistics in~$S=A+B$ is known and the probabilities
$\alpha$ and $\beta$ for a photon to be in volumes $A$ and $B$,
respectively, are given. What, then, is the probability distribution
$W(k,n-k)$ for $k$ photons to be in $A$ and $n-k$ photons to be
in~$B$? Both classical and quantum solutions to this problem are
discussed in this work. The results obtained for the quantum statistics,
as shown below, include a photon bunching effect that looks similar
to the quantum interference effect.

\subsection{\label{sub:Classical_Particles}A flux of classical particles}

Let us begin with the assumption that beam $S$ consists of classical
non-interacting particles that appear in the selected volume independently
of each other, so that there are no correlations between the particles.
In this case, statistics $p_{n}(S)$ in volume $S=A+B$ is related
to the relevant statistics in $A$ and $B$ by the classical equation
of probability convolution: 
\begin{equation}
p_{n}(A+B)=\sum_{k=0}^{n}p_{k}(A)p_{n-k}(B),\label{eq:General_Convolution_of_p}
\end{equation}
 Equation \eqref{eq:General_Convolution_of_p} is known to be applicable
to random processes in areas $A$ and $B$ if there are no correlations
between these random processes~\cite{Polya Distribution}. In addition,
\eqref{eq:General_Convolution_of_p} takes into account the conservation
of energy, or the number of particles $n=k+(n-k)$, if light beam
$S$ is separated into two parts $A$ and~$B$.

Classical non-interacting particles also obey Poisson statistics in
the arbitrary volume~$A$, so that 
\begin{equation}
p_{k}(A)=\dfrac{(wA)^{k}}{k\,!}\, e^{-wA},\label{eq:Poisson_Statistics}
\end{equation}
where $w$ is the average number of particles per unit volume. The
Poisson statistics is known to hold for photons in a coherent radiation
field~\cite{Mandel_Wolf}, \emph{i.\,e.} in the case of amplitude
stabilized laser radiation.

Substituting \eqref{eq:Poisson_Statistics} into \eqref{eq:General_Convolution_of_p}
yields 
\begin{equation}
\frac{(A+B)^{n}}{n\,!}=\sum_{k=0}^{n}\frac{A^{k}B^{n-k}}{k\,!(n-k)\,!}.\label{eq:Binom}
\end{equation}
This equation is an identity, which after multiplying both sides by~$n!$
turns into the binomial theorem. Therefore, the Poisson statistics
\eqref{eq:Poisson_Statistics} satisfies the probability convolution
equation~\eqref{eq:General_Convolution_of_p}.

All the above are well-known facts that are presented here for the
sake of convenience in comparing classical and quantum solutions.

\subsection{\label{sub:Quantum_Particles}A flux of quantum particles satisfying
Bose-Einstein statistics}

Now let beam $S$ be thermal radiation, so that the photons in beam
$S$ obey the BE statistics: 
\begin{equation}
p_{k}=\dfrac{w^{k}}{(1+w)^{k+1}},\label{eq:Bose_Statistics}
\end{equation}
where $w$ is the degeneracy parameter, or the average number of particles
per mode (i.\,e. in a coherence volume). Writing photon statistics
in this form, one assumes that the sampling volume in the statistics
measurement coincides with the coherence volume.

The BE~statistics in an arbitrary phase-space volume $A$ is given~by:
\begin{gather}
p_{k}(A)=\dfrac{A(A+1)...(A+k-1)}{k\,!}\,\dfrac{w^{k}}{(1+w)^{k+A}}.\label{eq:Bose_in_A}
\end{gather}
 This formula was derived by Leonard Mandel from combinatorial considerations~\cite{Mandel-Bose_in_any_volume}.
Actually, \eqref{eq:Bose_in_A} is a negative binomial distribution.
If $A=1$ then \eqref{eq:Bose_in_A} becomes the usual expression
for the BE statistics \eqref{eq:Bose_Statistics} in a single cell
of the phase-space.

Equation \eqref{eq:Bose_in_A} can be conveniently presented as 
\begin{equation}
p_{k}(A)=\dfrac{A^{\overline{k}}}{k\,!}\,\dfrac{w^{k}}{(1+w)^{k+A}},\label{eq:Bose_in_A_short}
\end{equation}
 where 
\begin{equation}
A^{\overline{k}}=A(A+1)...(A+k-1)\label{eq:Rising_Factorial}
\end{equation}
 is the rising factorial, or Pochhammer's symbol.

We saw above that the Poisson statistics satisfies the system of probability
convolution equations~\eqref{eq:General_Convolution_of_p}. Let us
now determine whether the BE statistics satisfies~\eqref{eq:General_Convolution_of_p}.
Substituting \eqref{eq:Bose_in_A_short} into \eqref{eq:General_Convolution_of_p}
after obvious reductions yields 
\begin{equation}
\dfrac{(A+B)^{\overline{n}}}{n\,!}=\sum_{k=0}^{n}\dfrac{A^{\overline{k}}}{k\,!}\,\dfrac{B^{\overline{n-k}}}{(n-k)\,!}.\label{eq:Vandermonde_identity}
\end{equation}
This is the well-known Vandermonde's identity~\cite{Concrete_Math},
which is a generalization of the binomial theorem \eqref{eq:Binom}
for rising factorials. A proof of this identity is given in Appendix~1.

Thus, the BE statistics \eqref{eq:Bose_in_A_short} also satisfies
the probability convolution equations~\eqref{eq:General_Convolution_of_p}.
This is because the BE statistics ignores intensity fluctuations in
the light beam, and hence does not take into account correlations
of photon numbers in volumes $A$ and $B$, which was proven, among
other sources, in~\cite{Scully}.

Along with the BE statistics, the Glauber's statistics \cite{Glauber}
for a homogeneously broadened spectral line also satisfies~\eqref{eq:General_Convolution_of_p},
which can be shown by direct verification (see Appendix~2).

Generally speaking, \emph{any} proper photon statistics \cite{note1} may either satisfy the probability convolution equation \eqref{eq:General_Convolution_of_p} or not. In the former case, photon statistics does \emph{not} take into account intensity fluctuations while in the latter case it may take into account the intensity fluctuations and related photon correlations. In this article, only the first type photon statistics is considered.

\section{\label{sec:G_B_D}Generalized Binomial Distribution}

Dividing the $n$-th equation in \eqref{eq:General_Convolution_of_p}
by $p_{n}(A+B)$ yields 
\begin{equation}
\sum_{k=0}^{n}W(k,n\!-\! k)=1,\label{eq:Normalization}
\end{equation}
where 
\begin{equation}
W(k,n\!-\! k)=\frac{p_{k}(A)\, p_{n-k}(B)}{p_{n}(A+B)}.\label{eq:GBD}
\end{equation}
The idea behind this expression is that $W(k,n-k)$ is the probability
that $k$ photons are in volume $A$ on condition that $n-k$ photons
are in~$B$. The denominator in \eqref{eq:GBD} guarantees that this
probability is correctly normalized according to~\eqref{eq:Normalization}.
In other words, eq.~\eqref{eq:GBD} gives the probability distribution
of photons among parts $A$ and $B$ of volume $S$ given it contains
$n$ photons. Equation~\eqref{eq:GBD} is valid for an arbitrary
photon statistics that satisfies~\eqref{eq:General_Convolution_of_p}.

\subsection{{Classical statistics}\label{sub:GBD_in_Classical_statistics}}

In the classical case, all the probabilities in \eqref{eq:GBD} are
given by Poisson statistics~\eqref{eq:Poisson_Statistics}. Substituting
\eqref{eq:Poisson_Statistics} into \eqref{eq:GBD} and taking into
account \eqref{eq:alpha_beta} gives 
\begin{equation}
W(k,n\negthinspace-\negthinspace k)=\dfrac{n\,!}{k\,!(n\negthinspace-\negthinspace k)\,!}\alpha^{k}\beta^{n-k}.\label{eq:Binomial_Distribution}
\end{equation}
This is the binomial distribution that occurs when a flux of classical
non-interacting particles is separated into two parts with probabilities
$\alpha$ and~$\beta$.

Obviously, should any \emph{nonpoissonian} statistics be substituted
into \eqref{eq:GBD} then the binomial distribution \eqref{eq:Binomial_Distribution}
would not be obtained. Therefore, the binomial distribution is valid
\emph{only} if particles in beam $S$ obey Poisson statistics. The
reverse statement is also true: let $X$ and $Y$ denote numbers of
photons in volumes $A$ and $B$, correspondingly. If $X$ and $Y$
are independent random variables and the conditional distribution
of $X$ given $X+Y$ is binomial then both $X$ and $Y$ should obey
Poisson statistics (for details see~\cite{Poisson_Char}). In other
words, the binomial distribution \eqref{eq:Binomial_Distribution}
and the Poisson statistics are closely interconnected phenomena --
in a beam of photons one is impossible without the other. For this
reason, it would be a mistake to use the binomial distribution with
any non-Poisson statistics. In the next Section the replacement for
the binomial distribution is found in the case of BE statistics.

Equation~\eqref{eq:GBD}, according to its derivation, holds for
an arbitrary photon statistics that is a solution of~\eqref{eq:General_Convolution_of_p}.
An arbitrary photon statistics inevitably tends towards Poisson statistics
in the limit of small photon density because, in this limit, the average
distance between photons is much greater than the coherence length
and, therefore, the photons may not take part in quantum interference
and should behave like classical noninteracting particles that obey
the Poisson statistics. Thus, for any photon statistics, Eq.~\eqref{eq:GBD}
will include the binomial distribution as a special case in the limit
of small photon density.

Consequently, equation \eqref{eq:GBD} can be regarded as a \emph{Generalized
Binomial Distribution}, which is valid for arbitrary statistics $p_{k}(A)$
that satisfies the probability convolution equations~\eqref{eq:General_Convolution_of_p}.

The binomial distribution \eqref{eq:Binomial_Distribution} is obtained
for Poisson statistics on the condition of uniformly distributed radiation
statistical properties over the beam cross-section, according to the
statement of the problem. For a nonuniform distribution of statistical
properties (for example, nonuniform distribution of radiation intensity)
the photon density $\omega$ will vary across the beam cross-section.
In this case, the probabilities $\alpha$ and $\beta$ for a photon
to be in $A$ and $B$, respectively, will be given by some integral
expressions rather than by~\eqref{eq:alpha_beta}. However, it can
be easily shown that this will not change the final result \eqref{eq:Binomial_Distribution}
that includes only the probabilities. This conclusion is in line with
the results presented in~\cite{Poisson_Char}.

\subsection{\label{sub:GBD_in_Quantum_Statistics}{Quantum statistics}}

In the quantum case, all the probabilities in \eqref{eq:GBD} are
given by Eq.~\eqref{eq:Bose_in_A_short} for the BE statistics in
an arbitrary volume. This is true because the BE statistics, as well
as the Poisson statistics, satisfies~\eqref{eq:General_Convolution_of_p},
as shown in Section~\ref{sub:Quantum_Particles}. Therefore, substituting
\eqref{eq:Bose_in_A_short} into \eqref{eq:GBD} yields 
\begin{multline}
W(k,n\!-\! k)\!=\dfrac{p_{k}(A)p_{n-k}(B)}{p_{n}(S)}=\dfrac{n\,!}{k\,!(n\!-\! k)!}\dfrac{A^{\overline{k}}B^{\overline{n-k}}}{S{\,}^{\overline{n}}}\\
=C_{n}^{k}\dfrac{A(A\!+\!1)\ldots(A\!+\! k\!-\!1)B(B\!+\!1)\ldots(B\!+\! n\!-\! k\!-\!1)}{S(S+1)\ldots(S+n-1)}.\label{eq:GBD_in_BE_stat}
\end{multline}
 This probability distribution is known as the Polya distribution.
Eq.~\eqref{eq:GBD_in_BE_stat} determines the probabilities of different
photon-number distributions ($k,\, n\!-\! k$) between volumes $A$
and $B$ if $n$ photons in $S=A+B$ obey the BE statistics. Therefore,
\eqref{eq:GBD_in_BE_stat} is a replacement for the classical binomial
distribution in the case where the incident photons obey quantum statistics.

It is known that the Polya distribution takes into account aftereffects
that are alien to the Bernoulli process~\cite{Polya Distribution}.
This property of the Polya distribution is shown below to produce
the photon bunching effect.

Given $A=\alpha S$ and $B=\beta S$, eq.~\eqref{eq:GBD_in_BE_stat}
may be written using probability $\alpha$ that a photon is in $A$
and probability $\beta$ that a photon is in~$B$: 
\begin{equation}
W(k,n\!-\! k)=\dfrac{(\alpha S)^{\overline{k}}}{k\,!}\,\dfrac{(\beta S)^{\overline{n-k}}}{(n\!-\! k)\,!}\,\dfrac{n\,!}{S\,^{\overline{n}}}.\label{eq:Polya_Distribution}
\end{equation}
This form of the generalized binomial distribution in BE statistics
will be convenient for further calculations.

If $S\rightarrow\infty$ then the Polya distribution \eqref{eq:Polya_Distribution}
becomes the classical binomial distribution:
\begin{flalign}
\lim_{S\rightarrow\infty}W(k,n-k) & =\lim_{S\rightarrow\infty}\dfrac{(\alpha S)^{\overline{k}}}{k\,!}\,\dfrac{(\beta S)^{\overline{n-k}}}{(n-k)\,!}\,\dfrac{n\,!}{S\,^{\overline{n}}}\nonumber \\
 & =\dbinom{n}{k}\alpha^{k}\beta^{n-k},\label{eq:Polya_for_large_S}
\end{flalign}
which means that the classical binomial distribution \eqref{eq:Binomial_Distribution}
is applicable not only in the case of Poisson statistics (as noted
above), but also in the case of BE statistics in the limit of low
particle density~$n\ll S$. In this limit the mean distance between
photons is much greater than the coherence length and photons behave
on average as independent classical particles that do not tend to
bunch. In the opposite limit as $S\rightarrow0$ photons show nonclassical
properties, which will be discussed below.

Information on the degeneracy parameter $w$ is missing from~\eqref{eq:Polya_Distribution},
which means that the probability distribution \eqref{eq:Polya_Distribution}
does not depend on the radiation temperature and the frequency range
selected to study photon statistics. However, probabilities $W(k,n-k)$
depend on the phase space volume $S$ occupied by the photons. That
is the fundamental difference between \eqref{eq:Polya_Distribution},
which is valid for the BE statistics, and the classical binomial distribution~\eqref{eq:Binomial_Distribution},
which is valid for Poisson statistics.

The Polya distribution \eqref{eq:Polya_Distribution} gives the exact
quantum statistical solution to the problem of interaction of an arbitrary
number of photons of thermal radiation occupying arbitrary phase-space
volume $S$, with a classical device that separates the photon flux
into two parts with known probabilities $\alpha={A}/{S}$ and $\beta={B}/{S}$.

In this section it was shown that, in the case of BE statistics, the
generalized binomial distribution~\eqref{eq:GBD} takes the form
of Polya distribution that is known to occur in random processes that
differ from the Bernoulli process~\cite{Polya Distribution}.

\section{\label{sec:GBD_in_BE}Photon bunching in Bose-Einstein statistics}

This Section presents several examples of the generalized binomial
distribution in the BE statistics. Probabilities $W(k,n\!-\! k)$
of different photon-number states $(k,\, n\!-\! k)$ for $n$ photons
in volume $S$ are calculated on the basis of~\eqref{eq:Polya_Distribution}
in some simple cases, the number of photons in $A$ being designated
as $k$ and the number of photons in $B$ being $n-k$.

\subsection{\label{sub:One-input-photon}One photon }

If one photon is in volume $S=A+B$ then the probabilities of photon-number
states (1,0) and (0,1) designating a photon in $A$ or in $B$, respectively,
are given by \global\long\def\arraystretch{2.5}
\begin{equation}
\begin{array}{ll}
W(1,0)=\dfrac{(\alpha S)^{\overline{1}}}{1\,!}\,\dfrac{(\beta S)^{\overline{0}}}{0\,!}\,\dfrac{1\,!}{S\,^{\overline{1}}} & =\;\alpha,\\
W(0,1)=\dfrac{(\alpha S)^{\overline{0}}}{0\,!}\,\dfrac{(\beta S)^{\overline{1}}}{1\,!}\,\dfrac{1\,!}{S\,^{\overline{1}}} & =\;\beta.
\end{array}\label{eq:One_photon}
\end{equation}
Probabilities \eqref{eq:One_photon} turn out to be independent of
volume~$S$. One photon is found in $A$ with probability~$\alpha$
or in $B$ with probability $\beta$ as it should be according to~\eqref{eq:alpha_beta}.

\subsection{\label{sub:Two-input-photons}Two photons }

If there are two photons in $S$ then the probabilities of different
photon-number distributions between volumes $A$ and $B$, according
to \eqref{eq:Polya_Distribution}, are:
\begin{equation}
\begin{array}{ll}
W(2,0)=\dfrac{\alpha(\alpha S+1)}{S+1} & \;\longrightarrow\alpha^{2},\\
W(1,1)=\quad\dfrac{2\alpha\beta S}{S+1} & \;\longrightarrow2\alpha\beta,\\
W(0,2)=\dfrac{\beta(\beta S+1)}{S+1} & \;\longrightarrow\beta^{2}.
\end{array}\label{eq:Two_Photons}
\end{equation}
In this case probabilities $W(k,n\!-\! k)$ depend on volume $S$
occupied by the photons. The limit values of corresponding probabilities
as $S\rightarrow\infty$ are shown on the right-hand side of~\eqref{eq:Two_Photons}.
These limit values coincide with the well known classical results
based on the binomial distribution as it should be according to~\eqref{eq:Polya_for_large_S}.

\begin{figure}[ht]
 \noindent \begin{centering}
\includegraphics[width=8cm]{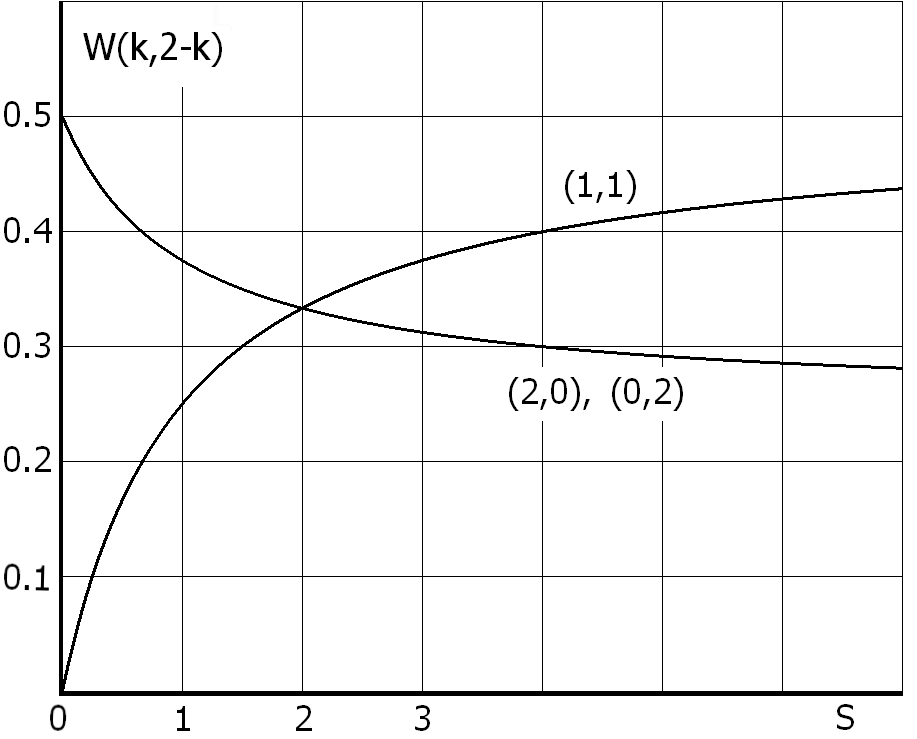} 
\par\end{centering}

\caption{{\small Non-classical probabilities of two-photon distribution among
two halves of volume~$S$ in the Bose-Einstein statistics. Volume
$S$ is given in the units of coherence volume.} {\label{fig:Two-photon-states}}}
\end{figure}

Probabilities \eqref{eq:Two_Photons} are shown in Fig.~\ref{fig:Two-photon-states}
as functions of volume $S$ occupied by the photons in the case where
$A=B$ ($\alpha=\beta=0.5$).

Decreasing volume $S$ makes classical probabilities invalid because
in quantum statistics, as follows from Fig.~\ref{fig:Two-photon-states},
photons tend to bunch together if they are located in a small phase-space
volume. Photon bunching is manifested by an abnormally high probability
of states (2,0) and (0,2) that designate both photons occupying only
one half of volume~$S$ if that volume becomes small, \emph{e.\,g}.
less than 2-3 coherence volumes. Note that $S=sc\tau$, therefore,
$S$ may be varied by changing the beam cross-section~$s$ and/or
sampling time~$\tau$.

\subsection{\label{sub:Three-input-photons}Three photons }

If three photons are in volume $S$ then using \eqref{eq:Polya_Distribution}
one obtains the following probabilities of different photon-number
configurations:
\begin{equation}
\begin{array}{ll}
W(3,0)=\alpha\dfrac{(\alpha S+1)(\alpha S+2)}{(S+1)(S+2)} & \;\longrightarrow\alpha^{3},\\
W(2,1)=3\alpha\beta\dfrac{S(\alpha S+1)}{(S+1)(S+2)} & \;\longrightarrow3\alpha^{2}\beta,\\
W(1,2)=3\alpha\beta\dfrac{S(\beta S+1)}{(S+1)(S+2)} & \;\longrightarrow3\alpha\beta^{2},\\
W(0,3)=\beta\dfrac{(\beta S+1)(\beta S+2)}{(S+1)(S+2)} & \;\longrightarrow\beta^{3}.
\end{array}\label{eq:Three_Photons}
\end{equation}

The basic features of the three-photon distribution among volumes
$A$ and $B$ are the same as in the previous example of two photons.
In the limit of large volume $S\rightarrow\infty$, shown on the right-hand
side of~\eqref{eq:Three_Photons}, photons behave as if they were
independent classical particles obeying the binomial distribution.
In contrast to such classical behavior, photons occupying a small
volume of about several bins exhibit a spatial distribution that deviates
markedly from the predictions of the classical binomial distribution.

\begin{figure}[ht]
\noindent \begin{centering}
\includegraphics[width=8cm]{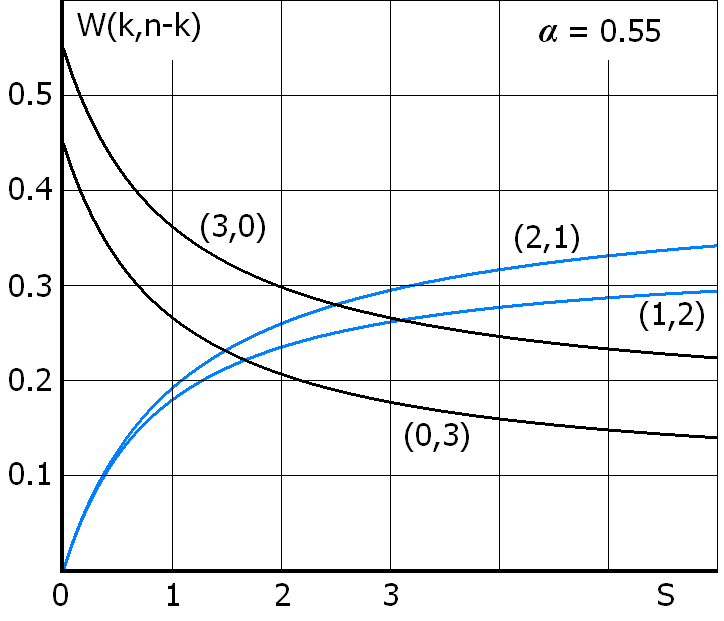} 
\par\end{centering}

\caption{{\small (Color online) Non-classical probabilities of three-photon
configurations versus volume $S$ occupied by the photons in a case
where the volume is divided into two unequal parts ($A=0.55\,S$).
The limit values of $W(k,3-k)$ for large $S$ coincide with classical
probabilities.} {\label{fig:Three-photon-states}}}
\end{figure}

Three-photon configuration probabilities \eqref{eq:Three_Photons}
are shown as functions of volume $S$ in Fig.~\ref{fig:Three-photon-states},
for a non-symmetrical division of $S$ into two parts ($\alpha=0.55$).
It is obvious from the plots that for small values of $S$, less than
3-4 coherence volumes, the probabilities $W(k,3\!-\! k)$ differ markedly
from the classical limits that are obtained as~$S\rightarrow\infty$.
This, again, is the manifestation of photon bunching in quantum statistics.

\subsection{\label{sub:Fifty-input-photons}Fifty photons}

\begin{figure}[ht]
\noindent \begin{centering}
\includegraphics[width=8cm]{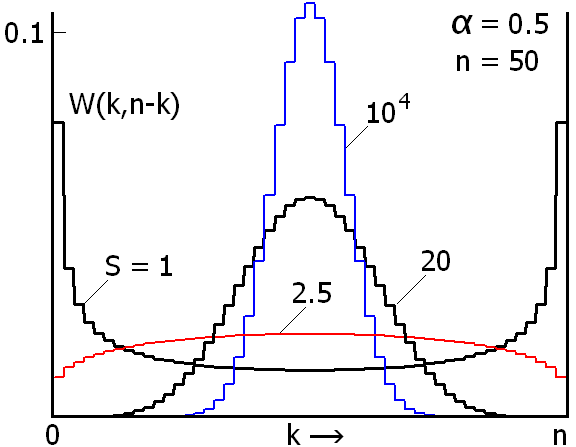} 
\par\end{centering}

\caption{{\small (Color online) Non-classical probability distributions of
50 photons among two equal parts of volume~$S$. Different curves
correspond to different phase-space volumes $S$ occupied by the photons
obeying the Bose-Einstein statistics. Histogram $S=10^{4}$ is close
to the classical binomial distribution while other histograms show
the deviation of quantum particles' distribution from the classical
distribution. Photon bunching, or a tendency toward coalescence, is
well manifested if photons are found in a small phase-space volume,
for example if~$S=1$. {\label{fig:50_photons}}}}
\end{figure}

As an example of bunching of a large number of photons in quantum
statistics, consider fifty photons in volume $S$ that is divided
into two equal parts ($\alpha=\beta=0.5$). Fig.~\ref{fig:50_photons}
presents probabilities $W(k,n-k)$ that $k$ photons are in part $A$
while $n-k$ photons are in part $B$ given total number of photons
in $S$ is $n=50$. Different curves correspond to different values
of phase-space volume $S$ occupied by the photons.

A case with $S=10^{4}$ is a good approximation to the limit of $S\rightarrow\infty$
because in this case each photon occupies, on average, a volume of
200 bins, and the probability distribution $W(k,n-k)$, according
to \eqref{eq:Polya_for_large_S}, looks like the classical binomial
distribution.

Fifty photons occupying a single bin (curve $S=1$) exhibit substantially
non-classical properties because the probability distribution in this
quantum state displays pronounced maxima at $k=0$ and $k=n$. This
is a manifestation of photon bunching, which is a tendency toward
collective behavior in quantum statistics. In this case the probability
that a group of photons is separated into two almost equal parts is
minimal, while in the classical case described by the binomial formula
this probability is maximal (curve $S=10^{4}$). In the limit $S\rightarrow0$,
a photon bunch looks like a single quantum entity that is not inclined
to dissolve into smaller groups of photons.

According to the above, in Poisson statistics (coherent radiation)
photons behave like classical particles in accordance with the classical
binomial distribution, while in quantum statistics (blackbody radiation)
photons exhibit different features showing a tendency toward bunching,
or collective behavior. In Fig.~\ref{fig:50_photons}, curve $S=10^{4}$
actually describes the behavior of classical particles while other
curves show the deviation of quantum particles from classical behavior
that becomes more pronounced for smaller phase-space volumes occupied
by the particles.

It is worthwhile noting that the average number of photons per one
cell in BE statistics is defined by the temperature and frequency
of radiation: 
\begin{equation}
w=\frac{1}{e^{\frac{h\nu}{kT}}-1}.\label{eq:w_in_BE}
\end{equation}
Therefore, 50 photons in one coherence volume could be observed with
notable probability if $kT\approx50h\nu$, \emph{i.\,e.} either in
a low-frequency range of blackbody radiation or for a black body of
very high temperature.

The Polya distribution \eqref{eq:GBD_in_BE_stat} and Fig.~\ref{fig:Two-photon-states}-\ref{fig:50_photons}
present the quantum solution to the problem stated in Section~\ref{sub:Statement}
regarding the distribution of $n$ photons among volumes $A$ and
$B$ in the case of BE statistics.

\section{\label{sec:Equivalence}Statistical equivalence of different problems}

The problem stated in Section~\ref{sub:Statement} is directly related
to the interaction of a light beam with a diaphragm because the distribution
$W(k,n-k)$ defines the probability that $k$ photons pass through
the hole $A$ while $n-k$ photons are absorbed by the screen~$B$
(see Fig.~\ref{fig:Problem_statement}). This problem is also related
to the following three problems:

1)~Photon number distribution at the output ports of a beamsplitter
of transmittance~$\alpha={A}/{S}$;

2)~Statistics of photon detection by a photodetector of quantum efficiency~$\alpha$.

3)~Photon statistics behind a neutral filter of transmittance~$\alpha$.

In all of the above problems, a photon flux is separated into two
parts with given probabilities $\alpha$ and $\beta=1-\alpha$. Although
very different physical processes are involved in these problems,
with regard to photon statistics they are equivalent and described
by the same formulas provided the light beam has the same properties
in all problems. This is because in statistics it is \emph{the probability
of an event that is important, not the nature of the event}.

For example, the photon statistics in part~$A$ of light beam~$S$
(Fig.~\ref{fig:Problem_statement}) should coincide with photon statistics
produced by the same light beam~$S$ after a beamsplitter of transmittance
$\alpha={A}/{S}$, because the probability that a photon is in volume~$A$
equals the probability that a photon is transmitted through the beamsplitter.
Likewise, if the probability of photodetection is~$\alpha$, the
same as the probability of a photon passing through the beamsplitter,
then the statistics of photo-absorptions should be the same as the
photon statistics behind the beamsplitter. A neutral filter actually
may be considered as a beamsplitter if the reflected radiation is
absorbed.

Consequently, the above problems are \emph{equivalent} with respect
to photon statistics, provided the light beams involved in these problems
have identical properties, \emph{i.\,e.} in the different problems
the incident photons are in the same quantum state. Therefore, any
statistical result obtained for one problem is sure to be valid for
the other equivalent problems as well. In particular, the Polya distribution
\eqref{eq:Polya_Distribution} and photon bunching effects shown in
Fig.~\ref{fig:Two-photon-states}-\ref{fig:50_photons} must hold
in all of the above problems if thermal photons are involved.

Given such an interpretation of the results, Fig.~\ref{fig:Two-photon-states}
shows the nonclassical probabilities of photon-number states at the
output ports of a beamsplitter versus the volume occupied by the two
photons before interacting with the beamsplitter, provided that the
photons obey the BE statistics and have entered the same input port
of the beamsplitter. Such photon behavior as that shown in Fig.~\ref{fig:Two-photon-states}
is not unusual: indeed, similar results were observed in a quantum
interference experiment \cite{Stability_of_clusters}, in which two
indistinguishable photons entered the \emph{same} input port of a
beamsplitter. The probability of output states (2,0) and (0,2) was
measured in this experiment against the distance between two input
photons and was found to be maximum for zero delay, just as indicated
in Fig.~\ref{fig:Two-photon-states} (for details refer to \cite{Stability_of_clusters}).
Regardless of the fact that non-thermal photons were used in this
experiment, the experimental results match well the curves presented
in Fig.~\ref{fig:Two-photon-states}. Qualitative coincidence of
experimental and theoretical curves indicates that the photon bunching
effect considered in this work does really exist and turns out to
be analogous to the quantum interference effect that was examined
in the above experiment.

The bunching of photons means that the photon statistics behind a
beamsplitter should, in a general case, deviate from the classical
binomial distribution, which holds only if the incident photons obey
the Poisson statistics. That conclusion was proven in deriving eq.~\eqref{eq:Binomial_Distribution}.
A similar result was obtained using a different theoretical method
in \cite{SVA-1,SVA-2} and later confirmed experimentally in \cite{Tim_Bartley_(2013)}
for a system of multiplexed on-off detectors.

The applicability of the results shown in Fig.~\ref{fig:Two-photon-states}
to the equivalent problems of photon interaction with a diaphragm,
photodetector, and neutral filter suggests that the quantum interference
phenomenon (or photon bunching effect) may also manifest itself in
situations where no beamsplitters are involved. This conclusion is
based on the fact that the photon bunching effect is a property of
photons in a certain quantum state rather than the property of a macroscopic
device.

\section{\label{sec:Conclusions}Conclusions}

The major results of this work are derived from two basic facts: firstly,
the probability convolution equations \eqref{eq:General_Convolution_of_p}
for the interaction of a light beam with a macroscopic device, 
\[
p_{n}(A+B)=\sum_{k=0}^{n}p_{k}(A)p_{n-k}(B)
\]
and, secondly, Mandel's formula \eqref{eq:Bose_in_A} for the Bose-Einstein
statistics in an arbitrary phase-space volume~$A$ 
\[
p_{k}(A)=\dfrac{A(A+1)...(A+k-1)}{k\,!}\,\dfrac{w^{k}}{(1+w)^{k+A}},
\]
 where $w$ is the degeneracy parameter of photon gas.

It is shown that the Mandel's formula satisfies the convolution equations,
which immediately yields the probability \eqref{eq:GBD_in_BE_stat}
of different photon-number states ($k,n-k$) after the interaction
of photon flux with a macroscopic device: 
\[
W(k,n-k)=\dfrac{p_{k}(A)p_{n-k}(B)}{p_{n}(S)}=C_{n}^{k}\dfrac{A^{\overline{k}}B^{\overline{n-k}}}{(A+B)^{\overline{n}}}.
\]
 The last formula is the Polya distribution that, according to the
analysis presented in Sections~\ref{sec:GBD_in_BE}-\ref{sec:Equivalence},
describes the photon bunching effect (which looks like quantum interference)
in the BE statistics.

Therefore, the main results of this work have been obtained as a direct
consequence of first principles, \emph{i.\,e.} the quantum statistics.
No additional assumptions were made and no approximate methods were
applied to obtain the results. For this reason, the Polya distribution
\eqref{eq:GBD_in_BE_stat} is the exact quantum statistical solution
of the stated problem.

The main conclusions of this work may be summarized as follows: 
\begin{enumerate}
\item It is shown that the Bose-Einstein statistics satisfies the system of probability
convolution equations~\eqref{eq:General_Convolution_of_p}, which is another proof of the well-established fact that the BE statistics does not take into account intensity
fluctuations and related photon correlations. Also, the fact that Mandel's formula \eqref{eq:Bose_in_A} is a solution of ~\eqref{eq:General_Convolution_of_p} implies that the Mandel's formula is valid for an arbitrary phase-space volume $A$ although originally it was obtained by Mandel \cite{Mandel-Bose_in_any_volume} only for a whole number of phase-space cells.
\item For probabilities of final photon-number states $(k,\, n)$ arising
if the photon flux is separated into two parts, a \emph{generalized
binomial distribution} is obtained 
\[
W(k,n)=\frac{p_{k}(A)\, p_{n}(B)}{p_{k+n}(A+B)}
\]
 that is applicable for arbitrary photon statistics $p_{k}(A)$ satisfying
the probability convolution equations. 
\item In the case of Bose-Einstein statistics, the generalized binomial
distribution takes the form of the Polya distribution, which presents
the exact quantum solution of the problem of interaction of an arbitrary
number of thermal photons occupying an arbitrary phase-space volume
with a macroscopic device that separates the photon flux into two
parts with known probabilities. 
\item Due to the statistical equivalence of different problems, the theoretical
results obtained for the interaction of photons of arbitrary statistics
with a diaphragm can be applied to the interaction of photons with
a beamsplitter, neutral filter, or photodetector. 
\item It is shown that the classical binomial distribution correctly describes
the probabilities of output photon-number states after the interaction
of photons with the macroscopic device in two cases only: (a) if the
photons obey the Poisson statistics; or (b) if the average distance
between photons greatly exceeds the coherence length (in this case
any photon statistics tends toward the Poisson statistics). Therefore,
these two conditions determine the domain of applicability of the
binomial formula. 
\item It follows from Fig.~\ref{fig:Two-photon-states}-\ref{fig:50_photons}
that photon bunching, or quantum interference in the BE statistics,
becomes notable if the mean occupation number is larger than unity~$\omega\gtrsim1$. 
\end{enumerate}

\section*{Acknowledgment}

The author would like to thank V.D. Ivanov and V.A. Kozminykh of Moscow
Institute for Physics and Technology, as well as A.V.~Masalov of
the Lebedev Physical Institute for stimulating discussions.

\section*{Appendix 1: Proof of identity (9)}
Equation \eqref{eq:Vandermonde_identity} includes the numerical sequence
\begin{equation}
f_{k}(A)=\dfrac{A^{\overline{k}}}{k\,!},\label{eq:f_k(A)}
\end{equation}
so that \eqref{eq:Vandermonde_identity} can be written as 
\begin{equation}
f_{n}(A+B)=\sum_{k=0}^{n}f_{k}(A)\, f_{n-k}(B)\label{eq:f_n(A+B)}
\end{equation}
The generating function of sequence \eqref{eq:f_k(A)}~is 
\begin{equation}
F(A)=\dfrac{1}{(1-z)^{A}},\label{eq:F(A)}
\end{equation}
which can be verified by direct expansion of \eqref{eq:F(A)} in a
Maclaurin series: 
\begin{equation}
\dfrac{1}{(1-z)^{A}}=\sum_{k=0}^{\infty}\dfrac{A^{\overline{k}}}{k\,!}z^{k}=\sum_{k=0}^{\infty}f_{k}(A)z^{k}.\label{eq:F(A)_expansion}
\end{equation}
Identity \eqref{eq:f_n(A+B)} is now validated by the following obvious
relation between the generating functions: 
\begin{equation}
\dfrac{1}{(1-z)^{A+B}}=\dfrac{1}{(1-z)^{A}}\,\dfrac{1}{(1-z)^{B}},\label{eq:Vandermonde_proof}
\end{equation}
 Q.E.D.

\section*{Appendix 2: The Glauber's statistics satisfies the probability convolution
equation}

Let $P(A)$ be the generating function (GF) of statistics $p_{k}(A)$
in an arbitrary phase-space volume~$A$. If this statistics satisfies
eq.~\eqref{eq:General_Convolution_of_p} then its GF obeys~\cite{Polya Distribution}
\begin{equation}
P(A+B)=P(A)P(B).\label{eq:P(A+B)}
\end{equation}
The functional equation \eqref{eq:P(A+B)} in $P$ has the solution
\begin{equation}
P(A)=P(1)^{A},\label{eq:Solution}
\end{equation}
where $P(1)$ is the GF of photon statistics in a unit volume $A=1$.
Obviously, any statistics satisfies \eqref{eq:General_Convolution_of_p}
if its GF satisfies~\eqref{eq:Solution}.

Recalling that by definition $A=ac\tau$, eq.~\eqref{eq:Solution}
may be written as 
\begin{equation}
P(A)=F(z)^{\tau},\label{eq:Solution_2}
\end{equation}
where $F(z)=P(1)^{ac}$.

The photon statistics obtained by Glauber~\cite{Glauber} for a Lorentzian
spectral line has GF (in Glauber's notations) 
\begin{equation}
Q(\lambda,\tau)=\exp\left\{ -\left[\left(\gamma^{2}+2\gamma W\lambda\right)^{{1}/{2}}-\gamma\right]\tau\right\} ,\label{eq:Glauber_GF}
\end{equation}
where $\gamma$ is the half-width of the spectral line, $W$ is the
average number of photons per second, $\tau$ is the sampling time
in the photon statistics measurement, and $\lambda=1-z$.

So, the GF of Glauber's statistics \eqref{eq:Glauber_GF} has the
form $Q(\lambda,\tau)=F(z)^{\tau}$, which coincides with~\eqref{eq:Solution_2}.
For this reason, Glauber's statistics satisfies~\eqref{eq:General_Convolution_of_p},
Q.E.D.


\begin{thebibliography}{10}
\bibitem[1]{Brown_Twiss}R. Hanbury Brown and R.Q. Twiss, Nature 177,
27 (1956)

\bibitem[2]{HOM_effect}C.\,K. Hong, Z.\,Y. Ou, and L. Mandel, ``\textit{Measurement of Subpicosecond Time Intervals between Two Photons by Interference},''
Phys. Rev. Lett. 59, 18, p.2045 (1987)

\bibitem[3]{Stability_of_clusters}G.\,Di Giuseppe, M. Atature, M.\,D.
Shaw, A.\,V. Sergienko, B.\,E.\,A. Saleh, M.\,C. Teich, A.\,J.
Miller, S.\,W. Nam, and J. Martinis, ``\textit{Direct observation of photon pairs at a single output port of a beam-splitter interferometer}'', Phys. Rev. A 68, 063817~(2003)

\bibitem[4]{Poisson_Char}S.\,D.\,Chatterji, ``\textit{Some elementary
characterizations of the Poisson distribution}'', Amer. Math. Monthly,
vol.\,70, pp.\,958-964 (1963)

\bibitem[5]{Polya Distribution}William Feller, ``\textit{An Introduction
to Probability Theory and Its Applications}'', vol. 1-2 (John Wiley \& Sons, 1970)

\bibitem[6]{Mandel_Wolf}L.\,Mandel and E.\,Wolf, ``\textit{Optical
Coherence and Quantum Optics}'' (Cambridge University Press, 1995)

\bibitem[7]{Mandel-Bose_in_any_volume}L.~Mandel, ``\textit{Fluctuations
of Photon Beams: The Distribution of the Photo-Electrons}'' Proc.
Phys. Soc. (London), 74, 233~(1959)

\bibitem[8]{Concrete_Math}Ronald L. Graham, Donald E. Knuth, Oren
Patashnik, ``\textit{Concrete Mathematics: A Foundation for Computer
Science}'', Second Edition (Addison-Wesley Publishing Company, Inc.,
1989)

\bibitem[9]{Scully}Marlan O. Scully and M. Suhail Zubairy, ``\textit{Quantum
optics}'' (Cambridge University Press, 2001)

\bibitem[10]{Glauber}Roy Glauber, \textit{Lecture \#17}, in C. DeWitt
(Ed.), ``Quantum Optics and Electronics'' (1965)

\bibitem[11]{SVA-1}J. Sperling, W. Vogel, and G. S. Agarwal, ``\textit{True
photo-counting statistics of multiple on/off detectors}'', Phys.
Rev. A 85, 023820 (2012)

\bibitem[12]{SVA-2}J. Sperling, W. Vogel, and G. S. Agarwal, ``\textit{Sub-binomial
light}'', Phys. Rev. Lett. 109, 093601 (2012)

\bibitem[13]{Tim_Bartley_(2013)}Tim J. Bartley, Gaia Donati, Xian-Min
Jin, Animesh Datta, Marco Barbieri, Ian A. Walmsley, ``\textit{Direct
observation of sub-binomial light}'', Phys. Rev. Lett.  110, 173602 (2013)

\bibitem[14]{note1}The term \textquotedbl{}photon statistics\textquotedbl{} is applicable only to random processes and does not apply to controlled processes, such as a light beam with given amplitude modulation, sub-Poissonian processes, etc., because in controlled processes photon statistics depends on the choice of starting points for the sampling intervals and, therefore, cannot be uniquely defined.

\end{thebibliography}
\end{document}